\documentclass[twocolumn]{aastex63}
\usepackage{graphicx}
\usepackage{float}
\usepackage{amsmath,amssymb}
\usepackage{xcolor}
\usepackage{multirow} 

\def\sn{SN\,2017ens}

\usepackage{colortbl}

\received{}
\revised{}
\accepted{}
\submitjournal{ApJ}

\shorttitle{Radio emission from SLSN\,2017ens}
\shortauthors{Margutti}
\graphicspath{{./}{figures/}}

\begin{document}

\title{Luminous Radio Emission from the Superluminous Supernova 2017ens at 3.3 years after explosion}

\correspondingauthor{Raffaella Margutti}
\email{rmargutti@berkeley.edu }

\author[0000-0003-4768-7586]{Raffaella Margutti}
\affiliation{Department of Astronomy, University of California, Berkeley, CA 94720-3411, USA}
\affiliation{Department of Physics, University of California, 366 Physics North MC 7300, Berkeley, CA 94720, USA}

\author[0000-0002-7735-5796]{J.~S. Bright}
\affiliation{Astrophysics, Department of Physics, University of Oxford, Keble Road, Oxford OX1 3RH, UK}
\affiliation{Department of Astronomy, University of California, Berkeley, CA 94720-3411, USA}

\author[0000-0002-4513-3849]{D.~J. Matthews}
\affiliation{Department of Astronomy, University of California, Berkeley, CA 94720-3411, USA}

\author[0000-0001-5126-6237]{D.~L. Coppejans} 
\affiliation{Department of Physics, University of Warwick, Coventry CV4 7AL, UK}

\author[0000-0002-8297-2473]{K.~D.~Alexander}
\affiliation{Department of Astronomy/Steward Observatory, 933 North Cherry Avenue, Rm. N204, Tucson, AZ 85721-0065, USA}

\author[0000-0002-9392-9681]{E.~Berger}
\affiliation{Center for Astrophysics \textbar{} Harvard \& Smithsonian, 60 Garden Street, Cambridge, MA 02138-1516, USA}

\author[0000-0002-0592-4152]{M. Bietenholz}
\affiliation{SARAO/Hartebeesthoek Radio Observatory, PO Box 443, Krugersdorp 1740, South Africa}

\author[0000-0002-7706-5668]{R. Chornock} 
\affiliation{Department of Astronomy, University of California, Berkeley, CA 94720-3411, USA}

\author[0000-0003-4587-2366]{L. DeMarchi}
\affiliation{Astronomer-in-Residence at Department of Physics Boise State University 1910 University Drive Boise, ID 83725-1570}

\author[0000-0001-7081-0082]{M.~R. Drout}
\affiliation{David A. Dunlap Department of Astronomy and Astrophysics, University of Toronto \\ 50 St. George Street, Toronto, Ontario, M5S 3H4 Canada}

\author[0000-0003-0307-9984]{T. Eftekhari}
\altaffiliation{NASA Einstein Fellow}
\affiliation{Center for Interdisciplinary Exploration and Research in Astrophysics (CIERA) and Department of Physics and Astronomy, Northwestern University, Evanston, IL 60208, USA}


\author[0000-0002-3934-2644]{W.~V.~Jacobson-Gal\'{a}n}
\altaffiliation{NSF Graduate Student Fellow}
\affiliation{Department of Astronomy, University of California, Berkeley, CA 94720-3411, USA}

\author[0000-0003-1792-2338]{T. Laskar}
\affiliation{Department of Physics \& Astronomy, University of Utah, Salt Lake City, UT 84112, USA}
\affiliation{Department of Astrophysics/IMAPP, Radboud University, P.O. Box 9010, 6500 GL, Nijmegen, The Netherlands}

\author[0000-0002-0763-3885]{D. Milisavljevic}
\affiliation{Purdue University, Department of Physics and Astronomy, 525 Northwestern Ave, West Lafayette, IN 47907 }
\affiliation{Integrative Data Science Initiative, Purdue University, West Lafayette, IN 47907, USA}

\author[0000-0002-5358-5642]{K. Murase}
\affiliation{Department of Physics, Department of Astronomy \& Astrophysics, \& Center for Multimessenger Astrophysics, Institute for Gravitation \& the Cosmos, The Pennsylvania State University, University Park, PA 16802, USA}
\affiliation{School of Natural Sciences, Institute for Advanced Study, Princeton, NJ 08540, USA}
\affiliation{Center for Gravitational Physics and Quantum Information, Yukawa Institute for Theoretical Physics, Kyoto, Kyoto 606-8502 Japan}

\author[0000-0002-2555-3192]{M. Nicholl}
\affiliation{Astrophysics Research Centre, School of Mathematics and Physics, Queens University Belfast, Belfast BT7 1NN, UK}

\author[0000-0002-9646-8710]{C.~M.~B. Omand}
\affiliation{The Oskar Klein Centre, Department of Astronomy, Stockholm University, AlbaNova, SE-106 91 Stockholm, Sweden}

\author[0000-0002-3019-4577]{M. Stroh} 
\affiliation{Center for Interdisciplinary Exploration and Research in Astrophysics (CIERA) and Department of Physics and Astronomy, Northwestern University, Evanston, IL 60208}

\author[0000-0003-0794-5982]{G. Terreran} 
\affiliation{Las Cumbres Observatory, 6740 Cortona Drive, Suite 102, Goleta, CA 93117-5575, USA}
\affiliation{Department of Physics, University of California, Santa Barbara, CA 93106-9530, USA}

\author{A.~Z. VanderLey} 





\begin{abstract}
We present the results from a multi-year radio campaign of the superluminous supernova (SLSN) \sn{}, which yielded the earliest radio detection of a SLSN to date at the age of  $\sim$3.3 years after explosion. \sn{} was not detected at radio frequencies in the first $\sim$300\,d of evolution but  reached $L_{\nu}\approx 10^{28}\,\rm{erg\,s^{-1}\,cm^{-2}}$ at $\nu\sim 6$\,GHz,  $\sim1250$ days post-explosion.  Interpreting the radio observations in the context of synchrotron radiation from the supernova shock interaction with the circumstellar medium (CSM), we infer an effective mass-loss rate $\dot M\approx 10^{-4}\,\rm{M_{\sun}\,yr^{-1}}$ at $r\sim 10^{17}$ cm from the explosion's site, for a wind speed of $v_w=50-60\,\rm{km\,s^{-1}}$ measured from optical spectra.  These findings are consistent with the spectroscopic metamorphosis of \sn{} from hydrogen-poor to hydrogen-rich $\sim190$\,d after explosion reported by \cite{Chen17ens}.  \sn{} is thus an addition to the sample of hydrogen-poor massive progenitors that explode shortly after having lost their hydrogen envelope. The inferred circumstellar densities, implying a CSM mass up to $\sim0.5\,\rm{M_{\sun}}$,  and low velocity of the ejection point at binary interactions (in the form of common envelope evolution and subsequent envelope ejection) playing a role in shaping the evolution of the stellar progenitors of SLSNe in the $\lesssim 500$ yr preceding core collapse. 
\end{abstract}

\keywords{(supernovae: 2017ens) }


\section{Introduction} \label{Sec:Intro} 

More than a decade after the identification of super-luminous supernovae (SLSNe) as a new class of stellar explosions with peak bolometric luminosities $L_{pk}\sim 10-100$ times those of ordinary core-collapse SNe \citep{Quimby11,Chomiuk11}, the nature of the energy source that powers their exceptional optical display and of their progenitor stars are still debated (see e.g., \citealt{Moriya18review,GalYam19ARA} for recent reviews). Arguments based on the comparison between the observed rise time of SLSNe and the diffusion time scale of photons through the explosion's ejecta lead to the conclusion  that the radioactive decay of large amounts (i.e.,  $> M_{\sun}$) of $^{56}$Ni that was suggested for example in the case of SN\,2007bi \citep{Gal-Yam09onSN2007bi} is not a viable option for the entire class of SLSNe, and that other sources of energy have to be invoked. 

Alternatives include (i) the SN shock interaction with a dense medium (e.g., \citealt{Chevalier11}) and/or (ii) a magnetar central engine (e.g., \citealt{Kasen10,Woosley10}). Both scenarios are expected to leave clear imprints on the non-thermal spectrum of the source, as opposed to the UV/optical/NIR regime that is dominated by thermal processes and is not sensitive to the explosion's fastest ejecta. Non-thermal processes, including the later-time break out of the emission from a pulsar wind nebula (e.g., \citealt{Omand23}),  are best constrained at radio frequencies. Yet, SLSNe have mostly eluded radio detection.  So far, PTF10hgi was the only SLSN with any detected radio emission, with the first detection only at $\delta t_{\rm{rest}}\sim6.3$ yr after explosion \citep{Eftekhari19,Eftekhari21, Law19, Mondal20, Hatsukade21PTF10hgivar,Hatsukade21sample}. 
Here we present the results from a multi-year radio campaign of \sn{}, which led to the earliest radio detection of a SLSN to date at $\delta t_{\rm rest} \sim 3.3$\,yr after explosion \citep{Coppejans2021GCN2017ens}\footnote{We note that further analysis and additional follow-up observations of the H-poor SLSN 2020tcw show that the radio emission that we reported in \cite{Coppejans21GCN2020tcw} is likely due to an unrelated source (Matthews et al., in prep.). }.

 \begin{figure}[!t]
\centering
    \includegraphics[width=0.45\textwidth]{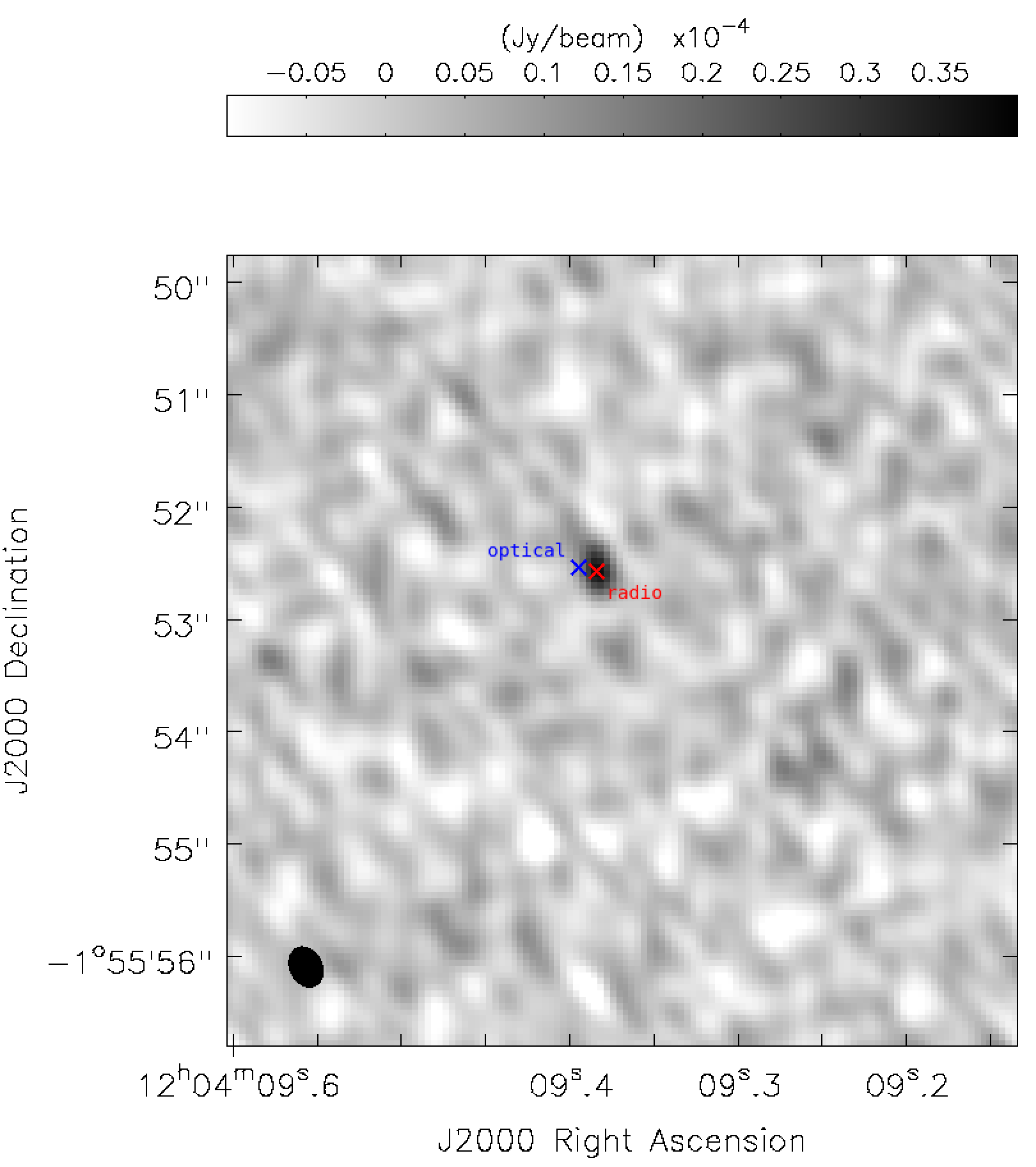}
     \caption{The field of \sn{} observed at 6\,GHz with the VLA on 2021 Jan 28 showing the onset of radio emission from the source at $F_{\nu}=38\pm5\,\mu$Jy. The gray-scale image shows flux density per beam and ranges between $-10$ and 40\,$\mu$Jy; the synthesized beam, with major and minor axes of 0.37\arcsec\ and 0.29\arcsec\ at a position angle of 26.6\degr, is shown in black in the bottom left corner of the image. The red cross shows the radio position, RA=$12^{\rm{h}}04^{\rm{m}}09^{\rm{s}}.384(1)$ Dec=$-01\degr55\arcmin52.56(2)\arcsec$,
     which is consistent with the optical position of the transient, shown by the blue cross, see \S\ref{Sec:Data}.}
 \label{Fig:DiscoveryImage}
 \end{figure}

\sn{} (aka ATLAS17gqa) was discovered by the Asteroid Terrestrial-impact Last Alert System (ATLAS; \citealt{tonry18}) on 2017 June 5, and classified as a H-poor stellar explosion at $z=0.1086$ by \cite{Chen17ens}. With a peak absolute magnitude $M_g=-21.1$\,mag, \sn{} belongs to the class of SLSNe. Starting with blue, featureless spectra until the  time of maximum light, \sn{} later underwent a dramatic spectral evolution characterized by the appearance of prominent H lines of the Balmer series that displayed a broad emission component (FWHM of $\sim$2000\,km\,s$^{-1}$) and a low-velocity P-Cygni profile with $v\sim 50-60\,\rm{km\, s^{-1}}$ \citep{Chen17ens}. Interestingly, \cite{Chen17ens} also report the presence of coronal lines likely resulting from X-ray photo-ionization that are typically seen for some type-IIn SNe (e.g., SN2010jl, \citealt{Fransson14}). These spectral features and the significant flattening of the optical light-curve at 
$\delta t_{\rm{rest}} >150$\,d since explosion (see Fig.\ 1 in \citealt{Chen17ens}) are a clear indication of the SN shock interaction with a dense, H-rich circum-stellar medium (CSM). The infrared (IR) brightening of \sn{} at a few hundred days post explosion reported by \cite{Sun2217ens} supports the presence of dust in the explosion's surroundings. While the origin of the dust is debated (newly-formed vs. pre-existing), IR excesses are routinely detected around SN shocks that interact with dense CSM (e.g., \citealt{Tinyanont19}). SN shocks propagating into a dense CSM are also well known particle accelerators and efficiently convert the explosion's kinetic energy into heat (e.g., \citealt{Chevalier17} for a recent review), a process that leads to copious X-ray and radio emission that we study here.  

Here we present radio observations of \sn{} spanning $\delta t = 55.0 - 1602$ d. This paper is organized as follows. In \S\ref{Sec:Data} we present the data anaylsis and reduction from our multi-year radio campaign of \sn{}. We model the radio data in \S\ref{Sec:RadioModelling} in the context of synchrotron emission from a blastwave propagating into the environment and we discuss our findings in \S\ref{Sec:Conc}. Following \cite{Chen17ens} we adopt $z=0.1086$, which translates into a luminosity distance of $d_L=490$\,Mpc for a cosmological model with $H_0=72\,\rm{km\,s^{-1}\,Mpc^{-1}}$, $\Omega_\Lambda=0.73$, $\Omega_m=0.27$. Times, $\delta t$,  are reported with respect to the explosion date, which is MJD $57907.8\pm 1.5$ \citep{Chen17ens}, and in the observer frame, unless explicitly noted otherwise. We note that the small uncertainty on the explosion date has no impact on our conclusions.

\section{Data analysis} \label{Sec:Data} 
 \begin{figure} 
 	\centering
 	\includegraphics[width=0.48\textwidth]{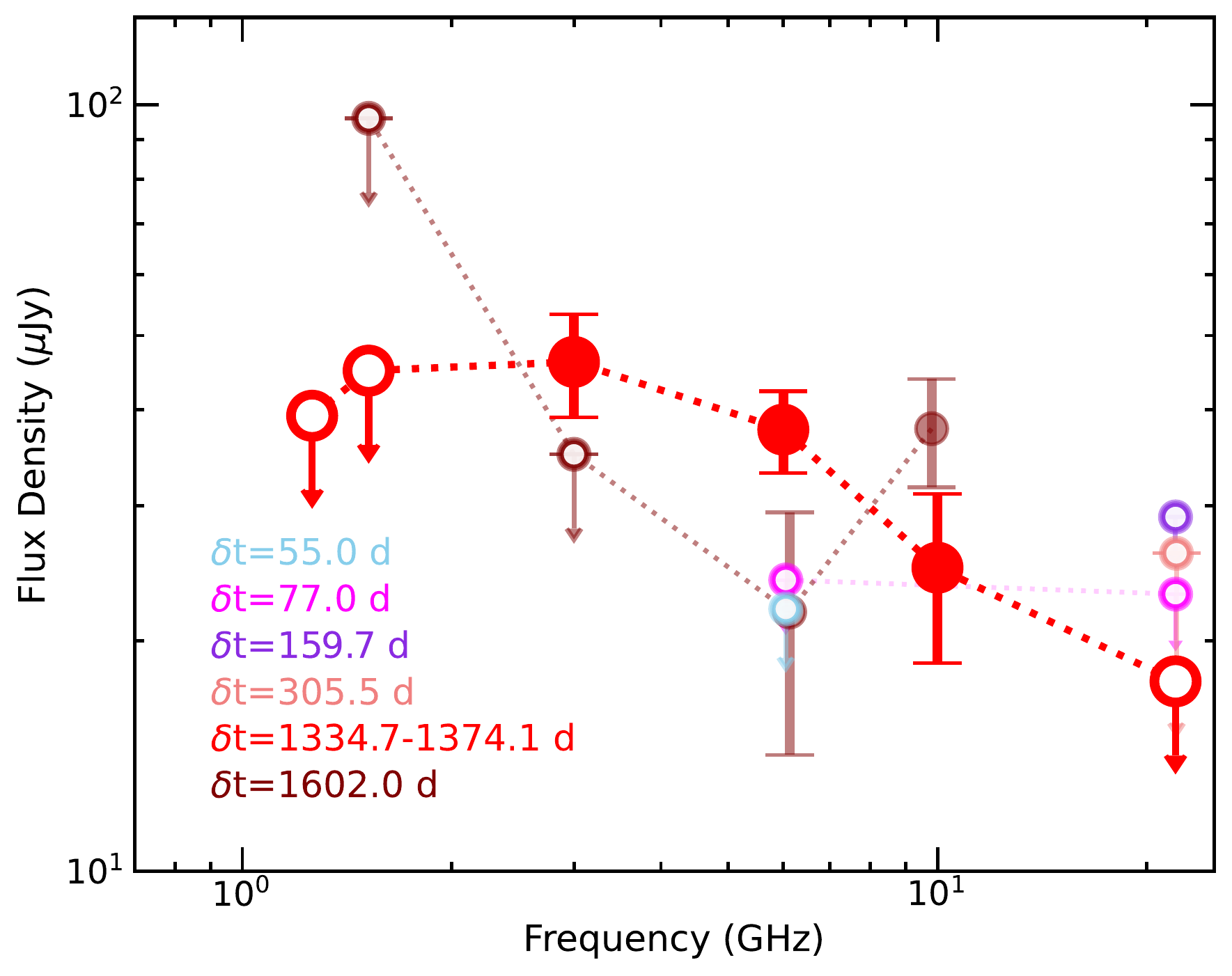}
     \caption{Temporal evolution of the radio spectral energy distribution (SED) of \sn{} in the time range $\delta t=55-1602$\,d. No radio emission is detected in the first year after explosion. 
     Our first radio detection of \sn{} occurred at $\delta t\sim1350$\,d (red). Empty symbols mark upper limits.}
 \label{Fig:radioSED}
 \end{figure}

 \begin{figure} 
  \hskip -0.5 cm
 	\includegraphics[width=0.53\textwidth]{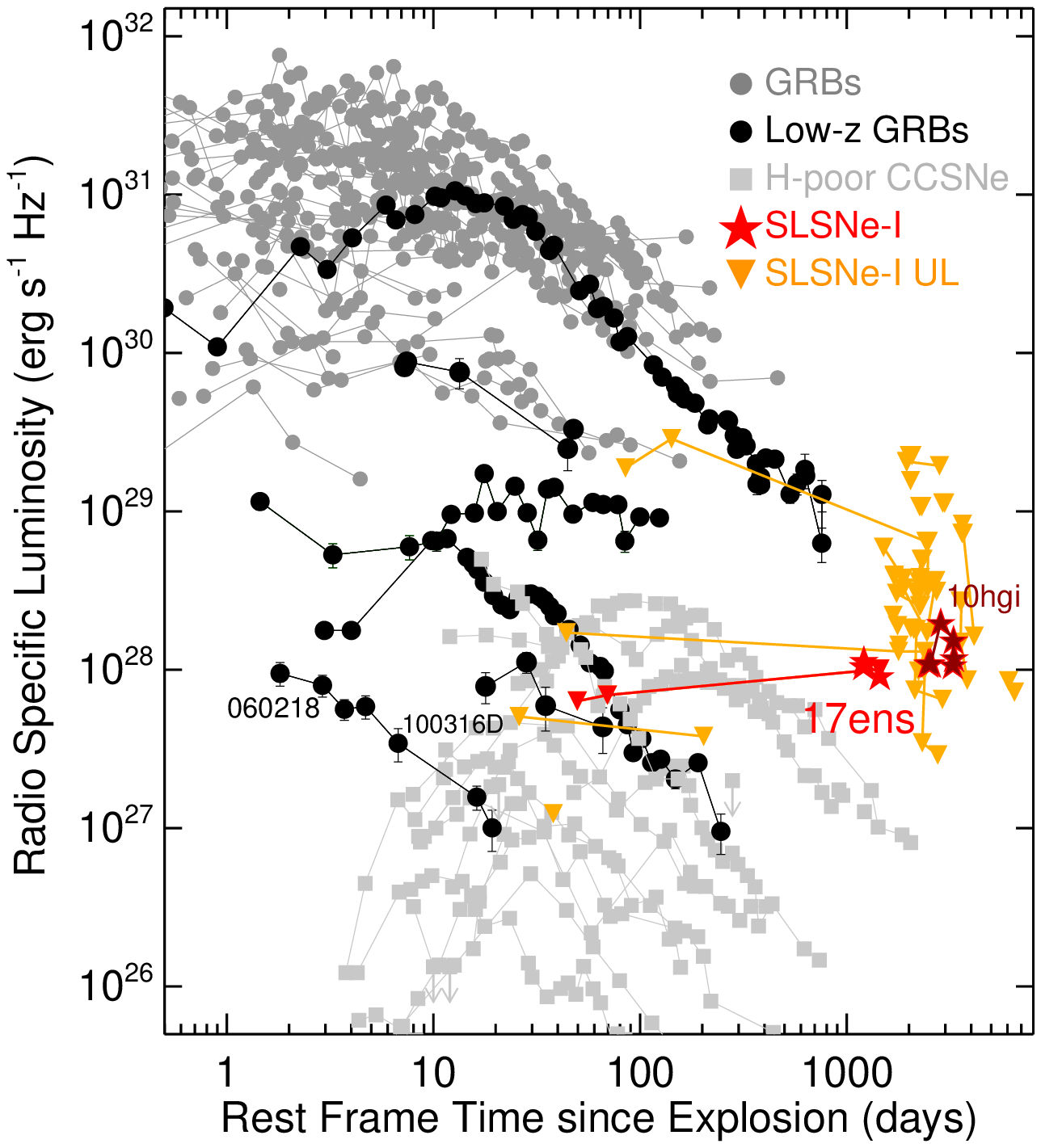}
     \caption{Radio specific luminosity $L_{\nu}$  from \sn{} in the context of other SLSNe-I (triangles and stars for upper limits and measurements, respectively), normal H-poor core-collapse SNe (light grey squares), long GRBs at cosmological distances (dark grey dots) and in the local universe (black dots) at $\nu\approx 6\,$GHz rest-frame. GRB and SN data collected from \cite{Soderberg10,Chandra12,Margutti14b} and references therein. For SLSNe-I, we select observations carried out in the rest-frame frequency range 3-8 GHz. SLSN-I references: \cite{Chandra09,Chandra10,Chomiuk11,Kasliwal16,Nicholl1615bn,Bright17egm,Bose18,Coppejans18,Hatsukade18,Nicholl1815bnlate,Schulze18,Eftekhari19,Law19,Mondal20,Chandra21,Eftekhari21,Hatsukade21PTF10hgivar}.}
 \label{Fig:RadioPhaseSpace}
 \end{figure}

We observed the field of \sn{} with the Karl G. Jansky Very Large Array (VLA) beginning on 2017 July 28 ($\delta t=55.0$\,d) as part of project 17A-480 (PI D. L. Coppejans). We continued observing \sn{} with VLA programs 17B-225 (PI R. Margutti), 20B-144 (PI D.~J. Matthews) and 21B-290 (PI D.~L. Coppejans). Overall, our multi-year radio monitoring spans the time period $\delta t=55.0-1602$\,d with the VLA in the A, B, and C configurations. We list the observing sessions in Table \ref{tab:radio}.

Data were calibrated using the Common Astronomy Software Applications (CASA; version 6.4.1.12; \citealt{McMullin07}) VLA pipeline version 2022.2.06.64, which performs flagging, delay correction, bandpass and absolute flux density scaling, and phase reference calibration. Imaging was performed either with the CASA \textsc{tclean} task, or WSClean \citep{offringa-wsclean-2017} using Briggs weighting with a robust factor of $0.5$.

We discovered significant radio emission from \sn{} on 2021 Jan 28 at $\delta t = 1334$ d (\citealt{Coppejans2021GCN2017ens}), with a flux density of $F_{\nu}=(38\pm5)\,\mu$Jy at 6 GHz (a $\sim7\,\sigma$ sigma detection given the image noise of $5\,\mu\rm{Jy/beam}$). We show the discovery image in Figure \ref{Fig:DiscoveryImage}. The position of the source is RA=$12^{\rm{h}}04^{\rm{m}}09^{\rm{s}}.384(1)$ Dec=$-01\degr55\arcmin52.56(2)\arcsec$, with the number in parenthesis indicating the uncertainty in the last digit.  We fit the flux density and position using the CASA task \texttt{IMFIT} by fitting to the image an elliptical Gaussian with 
dimensions fixed to those of the restoring (synthesized) beam, which has a major FWHM, minor FWHM, and position angle of 0.37\arcsec, 0.29\arcsec, and 26.6\degr, respectively. The absolute positional accuracy of the VLA using standard phase reference calibration techniques (as was the case for all our observations) and under typical conditions, is $\sim10\%$ of the synthesized beam, or $\sim0.03\arcsec$ to $\sim0.04\arcsec$ in this case. The optical position of \sn{} is RA=$12^{\rm{h}}04^{\rm{m}}09^{\rm{s}}.39(1)$ Dec=$-01\degr55\arcmin52.5(2)\arcsec$, with the number in parenthesis indicating the uncertainty in the last digit, as measured by the Gamma-Ray burst Optical Near-infrared Detector, GROND (J. Chen, priv. comm.), confirming the radio source is the counterpart to \sn{}. The optical source is also displaced compared to the host galaxy center (\citealt{Chen18}, their Fig. 4). 
For observations taken after our initial detection we fit a point source component fixed to the position found in our discovery image, and report the forced fit flux from this process. We present the results of our radio observations in in Table \ref{tab:radio}. In the following we also use a $3\sigma$ upper limit of $F_{\nu}<39.3\,\rm{\mu Jy}$ reported by \cite{Chandra21} using a Giant Metrewave Radio Telescope (GMRT) observation on 03.85 March 2021 (MJD 59281.85, $\delta t=1374.050$\,d) at 1.26 GHz.  
The complete set of radio observations is shown in Figure \ref{Fig:radioSED}. We put the radio lightcurve of \sn{} in the context of those of other core-collapse stellar explosions in Figure \ref{Fig:RadioPhaseSpace}.  

\section{Modelling of the Radio Emission} \label{Sec:RadioModelling} 
\begin{figure} 
  \hskip -1 cm
 	\includegraphics[width=0.53\textwidth]{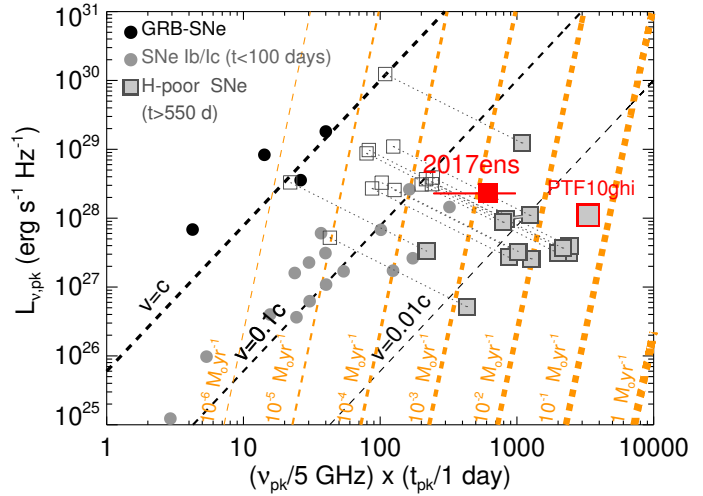}
     \caption{The radio SED at $\delta t=1334.7-1374.1\,\rm{d}$  constrains the location of \sn{} (red square) in the phase space of radio observables, spectral peak luminosity $L_{\nu, \rm{pk}}$ and ($\nu_{\rm{pk}} \times t_{\rm pk}$, where $\nu_{\rm{pk}}$ is the frequency of the spectral peak at time, $t_{\rm pk}$). Black (orange) dashed lines mark the location of constant outflow velocity (mass-loss rate $\dot M$, here given for $v_w=1000\,\rm{km\,s^{-1}}$). GRB-SNe (black circles) show ultrarelativistic outflows and low $\dot M$, while ordinary H-stripped SNe of type Ib/Ic at $t<100$\,d (grey circles) are associated with slower $v\sim0.1c$ shocks. The radio properties of \sn{} are similar to those of interacting H-stripped SNe at late times $t>550$\,d (grey squares, \citealt{Stroh21}, which were detected by the VLASS, \citealt{Lacy20}). If $\nu_{\rm{pk}}$ is not constrained we use an open square to mark the  SN location for an assumed optically thin spectrum $F_{\nu}\propto \nu ^{-1}$ and $\nu_{\rm{pk}}\approx0.3$ GHz as in \cite{Stroh21}. The location of the other H-poor SLSN with radio detection at the time of radio discovery (i.e. PTF10ghi, \citealt{Eftekhari19}) is marked. This plot assumes equipartition $\epsilon_e=\epsilon_B=1/3$.  References: \cite{Soderberg12,Stroh21}.  }
 \label{Fig:RadioSSA}
 \end{figure}

The emergence of detectable radio emission following the appearance of H lines in the spectrum suggests an origin of the radio emission connected with the interaction of the SN shock with a H-rich higher-density medium, in close similarity to other transitional events like SN\,2014C \citep{Milisavljevic15,Margutti17}. We explore in \S\ref{SubSec:radioenvironment} and \S\ref{SubSec:radioejet} the constraints on the CSM  density and on relativistic jets 
that can be placed with our radio observations.

\subsection{Constraints on the environment density}\label{SubSec:radioenvironment}
The deceleration of a SN shock in the CSM, and the resulting acceleration of particles at the SN shocks, is a well known source of radio synchrotron emission in stellar explosions (e.g., \citealt{Chevalier17}). In young SNe the radio emission originates from the forward shock and creates a characteristic bell-shaped spectrum with the frequency of the spectral peak cascading down to lower values with time as the emission becomes optically thin, in the case of type-Ib/c SNe usually to synchrotron self absorption (SSA). In the following we first assume that in the case of \sn{}, SSA dominates (i.e., that there is no significant free-free absorption), and adopt the standard formulation of SSA emission in SNe by \cite{Chevalier98} and follow the formalism that we have developed in \cite{deMarchi22}.\footnote{Specifically, we used Eqs.\ 18, 19, 21, 26, 27 from \cite{deMarchi22}.}

We start with considerations on the radio SED when we first detected the SN at $\delta t=1334.7-1374.1$\,d (Fig.\ \ref{Fig:radioSED}). If we define a power-law spectrum as one with $F_\nu \propto \nu^{-\beta}$, where $\beta$ is the  spectral index, then the observed values at $\delta t=1334.7-1374.1$\,d are not consistent with a single power-law, but rather imply $\beta  \approx 0.5-1$ for $\nu \gtrsim3.5$\,GHz,  with the upper 
limit at 22\,GHz requiring $\beta \gtrsim 0.5$, and the upper limits at $\nu < 2$\,GHz requiring a turnover somewhere below $\sim$4~GHz. The observations at this epoch constrain the observed spectral peak flux density\footnote{Note that this value is the actual peak in the spectrum. The corresponding peak flux parameter that would go into the \cite{Chevalier98} equations, which is the intersection of the optically thick and optically-thin asymptotes of the spectrum, is in the range 60-100 $\mu$Jy.} to $40\,\mu\rm{Jy} \lesssim F_{\rm{pk}}\lesssim 60\,\rm{\mu Jy}$ with the peak occurring in the range $1 \,\rm{GHz}\lesssim \nu_{\rm{pk}} \lesssim  4\,\rm{GHz}$.  Assuming equipartition of energy between electrons and magnetic field, with $\epsilon_e=\epsilon_B=0.33$, and a geometrical filling factor $f=0.5$, the observed $F_{\rm{pk}}$ and $\nu_{\rm{pk}}$ imply a forward-shock radius $R_{\rm{FS}}\approx(0.4-2.0)\times 10^{17}\,\rm{cm}$, and an average FS shock velocity $v_{\rm{FS}}/c \approx (0.01-0.06)$, which is $v_{\rm{FS}}\approx $ 3700-18800 $\rm{km\,s^{-1}}$.   The inferred post-shock magnetic field is $B=0.1-0.4\,\rm{G}$. An equivalent statement for these assumptions is that the CSM density at  $R_{\rm{FS}}$ corresponds to an effective $\dot M=(2-40)\times 10^{-4}\,\rm{M_{\sun}\,yr^{-1}}$ for $v_w=1000\,\rm{km\,s^{-1}}$, which is a density  $\rho_{\rm{CSM}}\approx 3\times 10^{-22}-1\times 10^{-19}\,\rm{g\,cm^{-3}}$ (particle density in the range $n=1.7\times 10^2-8.0\times 10^4\,\rm{cm^{-3}}$ for pure H composition).\footnote{We note that radio observations are sensitive to the CSM density $\rho_{\rm{CSM}}\propto \dot M/v_w$, which is why we report the assumed $v_w$ for  each of the inferred $\dot M$.}  The inferred shock internal energy is in the range $U_{eq}\approx (2-14)\times 10^{48}\,\rm{erg}$. 

Figure \ref{Fig:RadioSSA} shows the location of \sn{} at the time of this first radio detection in the $\nu_{\rm pk}$-$L_{\rm pk}$ plane, comparing it to the locations of other H-stripped core-collapse stellar explosions. \sn{} occupies a part of the parameter space that is populated by  SNe that showed late-time radio re-brightenings associated with shock interaction with a dense medium. For comparison purposes, we show  in Figure \ref{Fig:RadioSSA} the sample of late-time ``SN interactors'' that  were detected by  the Very Large Array Sky Survey (VLASS, \citealt{Lacy20}), as found by \citep{Stroh21}. In this context, the observed radio properties of \sn{} and the implied mass-loss rate are not unprecedented. We note however that the low-velocity P-Cygni profiles observed in the optical spectra of \sn{} indicate a H-rich wind velocity $v_w=50-60\,\rm{km\,s^{-1}}$ \citep{Chen17ens}, which is  significantly slower than the commonly assumed $v_w=1000\,\rm{km\,s^{-1}}$. The direct implication is that the effective progenitor mass-loss rate at $\sim 10^{17}\,\rm{cm}$ is $\dot M=(0.05-2)\times 10^{-4}\,\rm{M_{\sun}\,yr^{-1}}$.

Next we discuss the constraints on the innermost region of the CSM at distances $<10^{17}\,\rm{cm}$ that can be derived from the radio limits at $\delta t<1000$\,d. In a wind-like CSM density environment $\rho_{\rm{CSM}}\propto r^{-2}$ and constant microphysical parameters, the spectral peak frequency and flux density of an SSA spectrum are expected to evolve as $\nu_{\rm{pk}}\propto t^{-1}$ and $F_{\rm{pk}}\approx$ constant \citep{Chevalier98}. Assuming a $F_{\nu}\propto \nu^{5/2}$ optically-thick spectrum and a $F_{\nu}\propto \nu^{-1}$ optically-thin spectrum (as typically observed in SNe, e.g., \citealt{deMarchi22} and references therein), and extrapolating back in time the radio SED at $\delta t=1334.7-1374.1$\,d, we find that this model would 
violate the radio limits at $\delta t \sim 159.7-305.5$\,d. This finding  implies a deviation from a pure wind density profile (e.g.,  the presence of a shell of dense material encountered by the shock at $\approx R_{\rm{FS}}$, which is consistent with the delayed emergence of the H lines in the optical spectra and the flattening of the optical light curve) or significant free-free absorption (FFA), or both. Below we quantify the role of FFA at early times. 

We constrain the environment density at $r< 10^{17}\,\rm{cm}$ using radio observations at $\delta t\le 306$\,d. Self-consistently accounting for the possibility of external FFA in addition to SSA following \cite{weiler2002}, we find that the lack of detectable radio emission at $\delta t\le 306$\,d either implies a large free-free optical depth corresponding to $\dot M>2\times 10^{-2}\,\rm{M_{\odot}\,yr^{-1}}$ (for an assumed electron temperature $T_e=10^4$\,K)  or a shock propagation into a lower-density medium with $\dot M<6\times 10^{-5}\,\rm{M_{\odot}\,yr^{-1}}$ (all mass-loss rates quoted for $v_w=1000\,\rm{km\,s^{-1}}$). From the flux ratio of the narrow coronal lines [O III] $\lambda$4363 to $\lambda$5007 lines at $\approx 215$\,d,  \cite{Chen17ens} infer an electron number density in the CSM of $n_e\sim 10^{6}-10^{8}\,\rm{cm^{-3}}$ (for an electron temperature $T_e=50000-10000$\,K), which translates into an $\dot M\sim (0.002-0.2)\,\rm{M_{\sun}\,yr^{-1}}$ for $v_w=1000\,\rm{km\,s^{-1}}$ and a shock velocity $\gtrsim5000\,\rm{km\,s^{-1}}$. We  thus favour the high-$\dot M$ branch of the radio solution, which is $\dot M\approx 10^{-3}\,\rm{M_{\sun}\,yr^{-1}}$ for the more realistic $v_w=50-60\,\rm{km\,s^{-1}}$ measured from optical spectra.\footnote{We note that \cite{Chen17ens} also report a $\dot M\sim 4\times 10^{-4}\,\rm{M_{\sun}\,yr^{-1}}$ ($v_w=50\,\rm{km\,s^{-1}}$) from the modeling of the bolometric optical light-curve at $\delta t>150$\,d. This value is highly dependent on the assumed efficiency of conversion of kinetic energy into radiation, and it is consistent with the electron density that is inferred from the coronal line emission only for $r<10^{16}\,\rm{cm}$. Larger $\dot M$ would be required to meet the $n_e\sim 10^{6}-10^{8}\,\rm{cm^{-3}}$ constraint at  $r>10^{16}\,\rm{cm}$, which is consistent with our findings.}
We note that for these large densities, radio-emitting electrons and positrons may originate from inelastic $pp$ interactions, giving origin to synchrotron emission of secondary pairs from cosmic-ray ions, which is best revealed at mm wavelengths \citep{Murase14}. The late-time emergence of broad H spectral features  in \sn{}  still suggests a shell-like geometry of the H-rich CSM. For a CSM shell with $\Delta R \approx R_{\rm{FS}}$, our radio modelling indicates a total shell mass of $M_{\rm{CSM}}\approx 0.5\,\rm{M_{\sun}}$.

\subsection{Constraints on relativistic jets}\label{SubSec:radioejet}
\sn{} displayed broad spectral features with similarities to those observed in GRB-SNe \citep{Chen17ens}.  GRB-SNe are associated with relativistic jets with a variety of collimation and kinetic energy properties (e.g., \citealt{Hjorth12,Corsi21}) that manifest in the radio phase space as a diverse sample of radio light-curves spanning $\sim 4$ orders of magnitude in luminosity (Fig.\ \ref{Fig:RadioPhaseSpace}, black and grey filled circles). With reference to Fig.\ \ref{Fig:RadioPhaseSpace}, the  limits on the early radio emission of \sn{} at $\delta t<100$\,d clearly rule out on-axis jets of cosmological GRBs (grey filled circles) but leave the parameter space of the rapidly decaying radio emission associated with some low-luminosity GRBs unconstrained (e.g., GRBs 060218 and 100316D). 

Following \cite{Coppejans18}, we generated a grid of off-axis jet models using high-resolution, two-dimensional relativistic hydrodynamical jet simulations with Boxfit (v2, \citealt{vanEerten12}). The synchrotron radio emission originating from the deceleration of the jet in the environment depends on a set of intrinsic and extrinsic physical parameters. We explored isotropic-equivalent jet kinetic energy values in the range $ 10^{50} \le E_{\rm{k,iso}}\le 10^{55}$\,erg, medium densities $10^{-3}\le n \le10^2$\,$\rm{cm^{-3}}$ for an ISM-like density profile $\rho_{\rm{CSM}} \propto r^{0} $, and mass-loss rates $10^{-8}\le \dot M \le 10^{-3}\,\rm{M_{\sun}\,yr^{-1}}$ ($v_w=1000\,\rm{km\,s^{-1}}$) for a wind-like profile $\rho_{\rm{CSM}} \propto r^{-2} $. We selected jet half opening angles $\theta_{\rm jet}=[5\degr, 30\degr]$, as representative of a collimated and less-collimated outflow,  and observer angles $\theta_{\rm obs}=[30\degr, 45\degr,  90\degr]$.  For fiducial shock microphysical parameters $\epsilon_e=0.1$, $\epsilon_B=0.01$, $p=2.5$, our radio observations of \sn{} rule  out  jets with $\theta_{\rm jet}=5\degr$ ($\theta_{\rm jet}=30\degr$), $E_k\ge 10^{51}\,\rm{erg}$ expanding in an ISM-like medium with $n\ge 1\,\rm{cm^{-3}}$ ($n\ge 0.1\,\rm{cm^{-3}}$) for all observing angles. For a wind-like case the parts of the parameter space ruled out are $\dot M\ge 10^{-4}\,\rm{M_{\odot}\,yr^{-1}}$ ($\dot M\ge 10^{-5}\,\rm{M_{\odot}\,yr^{-1}}$),  $E_{k}\ge 10^{49.5}\,\rm{erg}$ ($E_{k}\ge 10^{50}\,\rm{erg}$) for $\theta_{\rm jet}=5\degr$ ($\theta_{\rm jet}=30\degr$) for all $\theta_{\rm obs}$. Free-free absorption has a minor impact on these conclusions, as most of the optically thick material is located in regions probed by the jet at times that are before our first epoch of observations.

\section{Discussion and Conclusions} \label{Sec:Conc}
We presented the earliest radio detection of a SLSN to date. 
The combination of the shallower optical light-curve decay and the late emergence of H emission in the optical spectra of a H-poor SN followed by detectable radio emission strongly suggests an origin of the radio emission related with the interaction of the explosion's shock  with a dense, H-rich medium.  Specifically, our radio analysis combined with inferences from high-resolution optical spectroscopy by  \cite{Chen17ens} suggests the presence of a dense shell of H-rich CSM. We connect the late emergence of the radio emission from \sn{} with a combination of the location of the H-rich material and an optical depth effect, i.e., with the time necessary for the radio emitting shock to reach a radius from which radiation could escape and reach the observer. Assuming spherical geometry, we estimate a CSM mass of $M_{\rm CSM}\lesssim 0.5\,\rm{M_{\sun}}$ within $\lesssim 10^{17}$\,cm.

\sn{} belongs to the small group of known H-poor SLSNe that developed H-emission in their spectra at $\delta t_{\rm rest}\ge100$\,d. This sample includes the SLSNe-I PTF10aagc, PTF10hgi, iPTF13ehe, iPTF15esb, iPTF16bad and SN\,2018bsz \citep{Yan15,yan17,Anderson18,Pursiainen22}. Similar to \sn{}, the phenomenology of these transitional SLSNe has been connected with the presence of dense, H-rich  CSM at distances of $\sim 10^{16}-10^{17}\,\rm{cm}$ from the explosion site  with estimated masses in the range $M_{\rm{CSM}}\sim 0.1-3\,\rm{M_{\sun}}$ (see e.g., Figures 12 and 13 from \citealt{Brethauer22}). These properties are not dissimilar to those inferred for transitional SNe that are \emph{not} of superluminous nature (i.e., the ``SN2014C-like'' events in \citealt{Brethauer22}), which suggests that the physical mechanism that drives the pre-SN mass ejections is independent from the superluminous nature of the stellar explosion. 

A key open question pertains to the physical origin of the H-rich CSM  mass, which is tied to the evolutionary path of the progenitor system in the final years before collapse. The slow CSM velocities $\sim50-60\,\rm{km\,s^{-1}}$ measured by \cite{Chen17ens} from optical spectra are not consistent with the significantly larger escape velocities ($\gtrsim 1000\,\rm{km\,s^{-1}}$) of material from compact H-deficient isolated massive progenitors like Wolf-Rayet stars. Instead these observations point at the envelope ejection of the primary exploding star as a result of binary interaction following a common envelope phase (e.g., \citealt{Podsiadlowski92}), as was proposed  for other transitional objects such as SN\,2014C and others (e.g., \citealt{Milisavljevic15,Sun2014C,Sun20}). 

The location of the H-rich CSM close to the explosion's site and the velocities measured imply an envelope ejection within $\approx 500$\,yr of core collapse. While the statistics are not complete, based on current optical spectroscopy observations, the fraction of SLSNe-I displaying a late-time emergence of H-features is of the order of $\lesssim 1/10$ (Blanchard, private communication). This fraction is broadly consistent with the expectations from a population of binary progenitor systems with a diverse set of initial properties and where only a small fraction of systems with wide orbital separations experience common envelope evolution and envelope ejection in the centuries before the collapse of the primary star. For example, \cite{Podsiadlowski92} estimate that $\sim6\%$ of binary systems with a primary star massive enough to explode as a SN experience ``case-C'' mass transfer (i.e., the progenitor fills its Roche lobe in a late evolutionary stage, which is associated with late-time envelope ejection). Updated binary synthesis simulations point at a fraction in the range $4\%-10\%$ of  progenitors of H-stripped SNe resulting from systems that experienced case-C common envelope evolution, with the range of values reflecting the assumed slope of the initial-mass function \citep{Margutti17}. This suggests  that (i) interacting binary systems might be common progenitors of  SLSNe-I; (ii) the physical ingredient that determines the superluminous nature of a SN (which constitute $\approx0.2\%$ of the core-collapse SN rate by volume in the local universe; \citealt{Li11,Quimby13}) is likely independent of the mechanism that leads to hydrogen envelope removal, which can instead operate in a wide variety of primary stars/progenitor systems. 

Finally, we address the implications of our deep constraints on the presence of relativistic jets in \sn{} in the broader context of radio observations of SLSNe-I. \sn{} and PTF10hgi  are the only two SLSNe-I for which radio emission has been detected (Figure \ref{Fig:RadioPhaseSpace}). The late-time 6\,GHz radio detection of PTF10hgi at $\delta t_{\rm{rest}}\approx 6.3$\,yr has been interpreted as emission from an off-axis jet \citep{Eftekhari19}, but the favored interpretation, based on the radio spectrum, is that of the emergence of emission from a pulsar wind nebula inflated by a magnetar central engine \citep{Eftekhari19,Law19,Mondal20}. At the time of writing, SN\,2011kl associated with GRB\,111209A is the only known example of stellar explosion that satisfies the superluminous criterion and that  has harboured a relativistic jet \citep{Greiner15}. Constraints on relativistic jets in a population of optically-discovered SLSNe-I have been presented in \citet{Coppejans18} and \citet{Eftekhari21}. The deep multi-year radio campaign on \sn{} provides additional insight. While low-energy outflows with $E_k\le 10^{50}$\,erg similar to those associated with low-luminosity GRBs (e.g., \citealt{Berger03b,Soderberg06c,Cano13,Margutti13b}) are not constrained, radio observations of \sn{} rule out energetic jets with properties similar to cosmological long GRBs expanding in environments typical of massive stars ($\dot M\ge 10^{-5}\,\rm{M_{\odot}\,yr^{-1}}$) for all observing angles, suggesting that either the CSM density along the jet direction is lower than around a typical massive star (and lower than our inferences of \S\ref{Sec:RadioModelling}), or that \sn{} did not launch such a jet. 

 The phase-space available to hide off-axis relativistic jets in SLSNe-I has significantly shrunk as a result of dedicated multi-year radio campaigns on the nearest SLSNe. Going forward, the very early identification of SLSNe, within days of explosion, will allow us to probe the weakest mildly relativistic jets (like those of GRBs 060218 and 100316D, Figure \ref{Fig:RadioPhaseSpace}). Our \sn{} effort highlights the scientific return of consistent and persistent radio monitoring of SLSNe from days until several years after the explosion, \emph{even in the case of radio non-detections in the first years}. Coupled with deep high-resolution optical spectroscopy that can reveal the chemical composition and dynamics of the CSM \citep{Chen17ens}, the radio emission from SN shocks provide a unique window into the final moments of evolution of massive stars that would not be otherwise accessible.

\section*{Acknowledgments}
The National Radio Astronomy Observatory is a facility of the National Science Foundation operated under cooperative agreement by Associated Universities, Inc.
The TReX team at UC Berkeley is supported in part by the National Science Foundation under Grant No. AST-2221789 and AST-2224255, and by the Heising-Simons Foundation under grant \# 2021-3248. 
T.E. is supported by NASA through the NASA Hubble Fellowship grant HST-HF2-51504.001-A awarded by the Space Telescope Science Institute, which is operated by the Association of Universities for Research in Astronomy, Inc., for NASA, under contract NAS5-26555. MRD acknowledges support from the NSERC through grant RGPIN-2019-06186, the Canada Research Chairs Program, the Canadian Institute for Advanced Research (CIFAR), and the Dunlap Institute at the University of Toronto.   MN is supported by the European Research Council (ERC) under the European Union’s Horizon 2020 research and innovation programme (grant agreement No.~948381) and by UK Space Agency Grant No.~ST/Y000692/1.

\appendix
\section{Radio Data Table}\label{AppendixRadio}

\startlongtable
\begin{deluxetable*}{ccccccccc}
\tablecaption{VLA observations of \sn{}}
\tablehead{
\colhead{Start Date} & \colhead{Project ID} & \colhead{VLA Config.} & \colhead{Centroid MJD} & \colhead{Phase$^{\rm{a}}$} & \colhead{Frequency} & \colhead{Bandwidth} & \colhead{Flux Density$^{\rm{b}}$} & RMS\\
 (dd/mm/yy) &  &  &  & (d) & (GHz) & (GHz) & ($\mu$Jy) & ($\mu$Jy/beam)
 }
\startdata
28/07/2017 & 17A-480 & C & 57962.839 & 55.039 & 6 & 2 & $<22$ & 7.3 \\ 
19/08/2017 & 17A-480 & C & 57984.785 & 76.985 & 22 & 8 & $<23$ & 7.6 \\ 
19/08/2017 & 17A-480 & C & 57984.818 & 77.018 & 6 & 2 & $<17$ & 5.7 \\
10/11/2017 & 17B-225 & B & 58067.521 & 159.721 & 22 & 8 & $<29$ & 9.7 \\ 
05/04/2018 & 17B-225 & A & 58213.270 & 305.470 & 22 & 8 & $<26$ & 8.7 \\ 
28/01/2021 & 20B-144 & A & 59242.481 & 1334.681 & 22 & 8 & $<18$ & 5.9 \\ 
28/01/2021 & 20B-144 & A & 59242.515 & 1334.715 & 6 & 4 & $38\pm5$ & 4.7 \\   
05/02/2021 & 20B-144 & A & 59250.306 & 1342.506 & 10 & 4 & $25\pm6$ & 6.5 \\    
05/02/2021 & 20B-144 & A & 59250.329 & 1342.529 & 3 & 2 & $46\pm7$ & 7.3 \\  
05/02/2021 & 20B-144 & A & 59250.352 & 1342.552 & 1.5 & 1 & $<45$ & 15 \\ 
22/10/2021 & 21B-290 & B & 59509.742 & 1601.942 & 10 & 4 & $38\pm6$ & 6.4 \\   
22/10/2021 & 21B-290 & B & 59509.759 & 1601.959 & 6 & 4 & $22\pm8$ & 7.8 \\   
22/10/2021 & 21B-290 & B & 59509.775 & 1601.975 & 3 & 2 & $<35$ & 11.6 \\ 
22/10/2021 & 21B-290 & B & 59509.791 & 1601.991 & 1.5 & 1 & $<96$ & 32.0 \\ 
\enddata
\tablecomments{$^{\rm{a}}$ Days since explosion (which is MJD  57907.8), using the midpoint time of the exposure on source. $^{\rm{b}}$ Uncertainties are 1$\sigma$, and upper limits are $3\sigma$. The listed uncertainties take a systematic uncertainty of 5\% into account. We fit a point source to the image (i.e., an elliptical Gaussian with the same dimensions as the restoring beam) to derive the flux density after our detection at C-band on 28/01/2021, and give the fitted flux density. If this is not formally a $3\sigma$ detection we give the $3\sigma$ upper limit.
\label{tab:radio}}
\end{deluxetable*}

\bibliography{BibliographyFile}{}
\bibliographystyle{aasjournal}

\end{document}